\documentclass[fleqn,usenatbib]{mnras}

\usepackage[T1]{fontenc}

\DeclareRobustCommand{\VAN}[3]{#2}
\let\VANthebibliography\thebibliography
\def\thebibliography{\DeclareRobustCommand{\VAN}[3]{##3}\VANthebibliography}

\usepackage{graphicx}	% Including figure files
\usepackage{amsmath}	% Advanced maths commands
\usepackage{amssymb}	% Extra maths symbols
\usepackage{newtxtext,newtxmath} %%%% BP: line has been added here
\usepackage{soul}
\usepackage{multirow}

\usepackage{lineno}

\usepackage{lscape}
\usepackage{longtable}
\usepackage{rotating}
\RequirePackage{rotating}
\begin{document}
%\linenumbers
% Please keep new commands to a minimum, and use \newcommand not \def to avoid
% overwriting existing commands. Example:
%\newcommand{\pcm}{\,cm$^{-2}$}	% per cm-squared

\newcommand{\micmap}[1]{\textcolor{magenta}{MM: #1}}

\newcommand{\barbara}[1]{{\em Barbara: }{\color{cyan}{#1}}}
\newcommand{\mg}[1]{\textcolor{green}{MG: #1}}
\newcommand{\pda}[1]{\textcolor{red}{PDA: #1}}
%%%%%%%%%%%%%%%%%%%%%%%%%%%%%%%%%%%%%%%%%%%%%%%%%%

%%%%%%%%%%%%%%%%%%% TITLE PAGE %%%%%%%%%%%%%%%%%%%

% Title of the paper, and the short title which is used in the headers.
% Keep the title short and informative.
%\title[Short title, max. 45 characters]{Prospects for multi-messenger detection of binary neutron star mergers in O4}
\title[Prospects for multi-messenger BNS  detections]{Prospects for multi-messenger detection of binary neutron star mergers in the fourth LIGO-Virgo-KAGRA observing run}

% The list of authors, and the short list which is used in the headers.
% If you need two or more lines of authors, add an extra line using \newauthor
\author[B. Patricelli et al.]{Barbara Patricelli,$^{1,2,3,4}$\thanks{E-mail: barbara.patricelli@pi.infn.it (BP); maria.bernardini@inaf.it (MGB)}
Maria Grazia Bernardini,$^5$\footnotemark[1]
Michela Mapelli,$^{6,7,8}$
Paolo D'Avanzo,$^5$ 
\newauthor
Filippo Santoliquido,$^{6,7}$ 
Giancarlo Cella,$^{3}$ 
Massimiliano Razzano$^{1,3}$ 
and Elena Cuoco$^{2,9,3}$\\
%%\micmap{io sono in favore di spellare i nomi propri per intero, allunga la author list ma così "il mondo" si ricorda di te}
%A. N. Other,$^{2}$
%Third Author$^{2,3}$
%and Fourth Author$^{3}$
%\\
% List of institutions
$^1$ Physics Department, University of Pisa, Largo B. Pontecorvo 3, I-56127 Pisa, Italy\\
$^{2}$ European Gravitational Observatory, Via E. Amaldi, I-56021 Cascina, Pisa, Italy\\
$^{3}$ INFN - Pisa, Largo B. Pontecorvo 3, I-56127 Pisa, Italy\\
$^{4}$ INAF - Osservatorio Astronomico di Roma, Via Frascati 33, I-00078 Monte Porzio Catone (Rome), Italy\\
$^{5}$ INAF - Osservatorio Astronomico di Brera, via Bianchi 46, I-23807 Merate (LC), Italy\\
$^{6}$ Physics and Astronomy Department Galileo Galilei, University of Padova, Vicolo dell’Osservatorio 3, I-35122, Padova, Italy\\
$^{7}$ INFN - Padova, Via Marzolo 8, I-35131 Padova, Italy\\
$^{8}$ INAF-Osservatorio Astronomico di Padova, Vicolo dell’Osservatorio 5, I-35122, Padova, Italy\\
$^9$ Scuola Normale Superiore (SNS), Piazza dei Cavalieri, 7 - I-56126 Pisa, Italy
}

% These dates will be filled out by the publisher
\date{Accepted XXX. Received YYY; in original form ZZZ}

% Enter the current year, for the copyright statements etc.
\pubyear{2022}

% Don't change these lines
%\begin{document}
\label{firstpage}
\pagerange{\pageref{firstpage}--\pageref{lastpage}}
\maketitle

\begin{abstract}
%It should be a single paragraph not more than 250 words (200 words for Letters).
The joint detection of GW170817 and GRB 170817A opened the era of  multi-messenger astronomy with gravitational waves (GWs) and provided the first direct probe that at least some binary neutron star (BNS) mergers are progenitors of short gamma-ray bursts (S-GRBs). In the next years, we expect to have more multi-messenger detections of BNS mergers, thanks to the increasing sensitivity of GW detectors. Here,  we present a comprehensive study on the prospects for joint GW and electromagnetic  observations of merging BNSs in the fourth LIGO--Virgo--KAGRA observing run with \emph{Fermi}, \emph{Swift}, INTEGRAL and SVOM. This work combines accurate population synthesis models with simulations of the expected GW signals and the associated S-GRBs, considering different assumptions about the GRB jet structure. We show that the expected rate of joint GW and electromagnetic detections could be up to $\sim$ 6 yr$^{-1}$ when \emph{Fermi}/GBM is considered. Future joint observations will help us to better constrain the association between BNS mergers and S-GRBs, as well as the geometry of the GRB jets.
\end{abstract}

\begin{keywords}
%keyword1 -- keyword2 -- keyword3
neutron star mergers -- gravitational waves  -- gamma-ray bursts %%\micmap{I don't think neutron star mergers is a  MNRAS keyword }
\end{keywords}

%%%%%%%%%%%%%%%%%%%%%%%%%%%%%%%%%%%%%%%%%%%%%%%%%%

%%%%%%%%%%%%%%%%% BODY OF PAPER %%%%%%%%%%%%%%%%%%

\section{Introduction}
On August 17th, 2017, Advanced LIGO \citep{2015CQGra..32g4001L} and Advanced Virgo \citep{2015CQGra..32b4001A} observed for the first time a gravitational wave (GW) signal from the inspiral of a binary neutron star (BNS) merger (GW170817, \citealp{2017PhRvL.119p1101A}); less than 2 seconds after, a short gamma-ray burst (S-GRB), GRB 170817A, was observed by \emph{Fermi}/GBM \citep{2017ApJ...848L..14G} and INTEGRAL \citep{2017ApJ...848L..15S}. Besides the discovery of GRB afterglow in the X-rays and radio band \citep{2017Natur.551...71T,2017Sci...358.1579H}, the electromagnetic follow-up of GW\,170817 also led to the discovery and characterisation of the bright kilonova AT2017gfo \citep{2017Natur.551...67P,2017Natur.551...75S,2017ApJ...848L..27T,2017Sci...358.1556C}. This joint detection marked the beginning of multi-messenger astronomy with GWs and provided the first direct evidence that BNS mergers are progenitors of S-GRBs \citep{2017ApJ...848L..13A}. The joint observation of GW170817 and GRB 170817A also allowed astrophysicists  to infer some basic properties of S-GRB jets. For instance, long baseline interferometry observations put constraints on the source size and its displacement, that were found to be consistent with expectations for a structured, relativistic jet \citep{2018Natur.561..355M,2019Sci...363..968G}. 
During the first half of their third observing run (O3a), Advanced LIGO and Advanced Virgo observed another potential BNS merging system: GW190425 \citep{2020ApJ...892L...3A}, but no electromagnetic (EM) signal  was found in association with this event (see, e.g., \citealp{2019ApJ...880L...4H}).  However, it would have been difficult to detect an EM counterpart to GW190425, given its poor sky localisation and larger distance compared with  GW170817. 

%In the next years, current generation GW detectors will operate with increased sensitivity and we expect more BNS mergers to be observed  \citep{2020LRR....23....3A}. At the same time, new EM detectors such as the Sino-French space mission SVOM (Space-based multi-band astronomical Variable Objects Monitor, \citealp{2016arXiv161006892W}) will start taking data, in synergy with GW detectors and other satellites such as the \emph{Fermi Gamma-ray Space Telescope} \citep{2007AIPC..921....3R}, \emph{INTEGRAL} \citep{2003A&A...411L...1W} and the \emph{Neil Gehrels Swift Observatory} \citep{2004ApJ...611.1005G}: multi-messenger astronomy will be the key to further probe the physics and astrophysics of BNS systems. 

Here, we present the prospects for joint GW and EM detection of BNS merging systems in the next observing run (O4) of the GW detector network \citep{2020LRR....23....3A} composed of  Advanced LIGO, Advanced Virgo and KAGRA \citep{2019CQGra..36p5008A}, with a particular focus on the prospects for the joint detection of a GW signal and of the emission of a S-GRB. In fact, during O4 the GW interferometers will operate with increased sensitivity, and the synergy with the facilities that are capable to trigger, locate and rapidly deliver the main characteristics of the S-GRB in nearly real time will be crucial for the unambiguous identification of the EM counterpart of the GW events, and for the follow-up at different wavelengths. These facilities such as the \emph{Fermi Gamma-ray Space Telescope} \citep{2007AIPC..921....3R}, INTEGRAL \citep{2003A&A...411L...1W} and the \emph{Neil Gehrels Swift Observatory} \citep{2004ApJ...611.1005G} will also be complemented by new EM detectors such as the Sino-French space mission SVOM (Space-based multi-band astronomical Variable Objects Monitor, \citealp{2016arXiv161006892W}), that will start taking data during the O4 observing run (mid-2023). 
This unique combination of GW and EM facilities will be the key to further probe the physics and astrophysics of BNS systems.

Our investigation combines accurate population synthesis models with simulations of the expected GW signals. The associated S-GRB emission is calculated with general assumptions on the S-GRB population and testing different possible models for the jet structure to include also slightly off-axis events. This paper is organized as follows. In Sec.  \ref{sec:population}, we describe the theoretical models used to simulate the population of BNS merging systems. Secs. \ref{sec:GW} and \ref{sec:GRB} present our simulation and analysis pipelines for GW signals and S-GRBs. In Sec. \ref{sec:results} we present our results. Finally, Sec. \ref{sec:conclusions} discusses  our results and summarizes our conclusions.

\section{The BNS population} \label{sec:population}

We generate a sample of synthetic BNSs  populating the local Universe up to a redshift $z=0.11$, that is consistent with the expected horizon\footnote{The horizon  is the farthest distance at which a source with optimal sky location and binary inclination can be detected above a threshold signal-to-noise ratio (SNR), generally defined as SNR=8.} for BNS mergers of Advanced Virgo and Advanced LIGO in the O4 configuration \citep{2020LRR....23....3A}. %To do this, we considered the cosmic BNS merger rate density as a function of redshift as estimated in  \cite{2021MNRAS.502.4877S}...

We obtained our catalogs of synthetic BNSs with the {\sc mobse} population-synthesis code \citep[the acronym {\sc mobse} stands for massive objects in binary stellar evolution, see, e.g., ][]{mapelli2017,giacobbo2018a,giacobbo2018b}. {\sc mobse} is an upgraded version of the code {\sc bse} \citep[binary stellar evolution, see][]{hurley2002}, including an up-to-date formalism for the winds of massive hot stars (O-type, B-type, Wolf-Rayet and luminous blue variable stars), which describes the mass loss rate as
$\dot{M}\propto{}Z^\beta{}$ for all massive hot stars, where
\begin{eqnarray}\label{eq:beta}
\beta{}=\left\{  
\begin{array}{ll}
0.85 & {\rm if}\quad{} \Gamma_e<2/3,\\
2.45-\Gamma_e\,{}2.4 & {\rm if}\quad{} 2/3\leq{}\Gamma_e\leq1,\\ 0.05 &  {\rm if}\quad{}\Gamma_e>1 
    \end{array}
    \right.
\end{eqnarray}
In equation~\ref{eq:beta}, $\Gamma_e$ is the Eddington ratio for electron scattering (see \citealt{giacobbo2018a} for details).

For the outcome of core-collapse supernovae we use the delayed model discussed by \cite{fryer2012}, which does not produce any mass gap between neutron stars (NSs) and black holes. Furthermore, {\sc mobse} treats electron-capture  and (pulsational) pair instability supernovae as described in \cite{giacobbo2019} and \cite{mapelli2020}, respectively. We model natal kicks as
\begin{equation}\label{eq:kick}
v_{\rm kick}=f_{\rm H05}\,{}\frac{m_{\rm ej}}{\langle{}m_{\rm ej}\rangle{}}\,{}\frac{\langle{}m_{\rm NS}\rangle{}}{m_{\rm rem}},
    \end{equation}
where $f_{\rm H05}$ is a random number extracted from a Maxwellian distribution with one-dimensional root-mean-square $\sigma_{\rm kick} = 265$ km~s$^{-1}$ \citep{hobbs2005}, $m_{\rm rem}$ is the mass of the compact remnant (neutron star or black hole), $m_{\rm ej}$ is the mass of the ejecta, while $\langle {m}_{\mathrm{NS}}\rangle $ is the average NS  mass, and $\langle {m}_{\mathrm{ej}}\rangle $ is the average mass of the ejecta associated with the formation of a NS of mass $\langle {m}_{\mathrm{NS}}\rangle $ from single stellar evolution. In our calculations, we adopt $\langle {m}_{\mathrm{NS}}\rangle =1.3$ $\,{M}_{\odot }$ and $\langle {m}_{\mathrm{ej}}\rangle =9$ $\,{M}_{\odot }$, respectively. As discussed in \cite{giacobbo2020}, equation~\ref{eq:kick} can be used for both NSs and black holes, produces lower kicks for ultra-stripped supernovae \citep{tauris2013,tauris2015,tauris2017}, matches the distribution of proper motions of young pulsars in the Milky Way \citep{hobbs2005} and the merger rate density inferred by the LIGO--Virgo collaboration \citep{abbottO3apopandrate}. Binary evolution has been implemented in {\sc mobse} adopting the formalism by \cite{hurley2002}. 

For this work, we ran three sets of simulations with {\sc mobse}, corresponding to three different choices of the common-envelope parameter $\alpha{}=1$, 3 and 7 \citep{hurley2002}. The $\alpha$ parameter encodes the efficiency of common envelope ejection: a large value of $\alpha$ means that the envelope can be efficiently removed from the binary system \citep{webbink1984}. Here, we choose these three values of $\alpha$, because they approximately bracket current uncertainties on the merger rate (see the discussion below). In all our models, we assume that main sequence and Hertzsprung gap stars cannot survive a common envelope phase. For the other stellar types, we calculate common envelope as described in \cite{hurley2002}. Each simulation set is composed of 12 metallicities $Z= 0.0002,$ 0.0004, 0.0008, 0.0012, 0.0016, 0.002, 0.004, 0.006, 0.008, 0.012, 0.016, and 0.02. We  simulated $10^7$ binary stars per each metallicity. Primary masses follow a Kroupa initial mass function \citep{kroupa2001} between 5 and 150 M$_\odot$. The values of orbital period, eccentricity and mass ratio are drawn from \citep{sana2012}. In particular, we derive the mass ratio $q = m_2/m_1$ as $\mathcal{F}(q)\propto{}q^{-0.1}$ with $q \in [0.1,\,{}1]$, the orbital period $P$ from $\mathcal{F}(\Pi)\propto{}\Pi^{-0.55}$ with $\Pi = \log_{10}{(P/{\rm day})} \in [0.15,\,{}5.5]$ and the eccentricity $e$ from $\mathcal{F}(e)\propto{}e^{-0.42}$  with  $0\leq{}e\leq{}0.9$.

From these runs, we derived catalogs of BNS masses and delay times, i.e. the time elapsed from the formation of a binary star to its merger. In order to derive the redshift of each merger, we fed these quantities to our code {\sc cosmo$\mathcal{R}$ate} \citep{santoliquido2020}. {\sc cosmo$\mathcal{R}$ate} calculates the BNS merger rate density evolution in the comoving frame  and then associates  a redshift of formation and merger to the simulated BNSs, as described in \cite{santoliquido2021}. In particular, {\sc cosmo$\mathcal{R}$ate} estimates the cosmic merger rate density $\mathcal{R}(z)$ in the comoving frame as
\begin{equation}
\label{eq:mrd}
   \mathcal{R}(z) = \frac{{\rm d}~~~~~}{{\rm d}t(z)}\left[\int_{z_{\rm max}}^{z}\psi(z')\,{}\frac{{\rm d}t(z')}{{\rm d}z'}\,{}{\rm d}z'\int_{Z_{\rm min}}^{Z_{\rm max}}\eta(Z) \,{}\mathcal{F}(z',z, Z)\,{}{\rm d}Z\right],
\end{equation}
    where $t(z)$ is the comoving time at redshift $z$, $Z_{\rm min}$ and $Z_{\rm max}$ are the minimum and maximum metallicity, $\psi{}(z')$ is the cosmic star formation rate density at redshift $z'$, $\mathcal{F}(z',z,Z)$ is the fraction  of compact binaries that form at redshift $z'$ from stars with metallicity $Z$ and merge at redshift $z$, and $\eta(Z)$ is the merger efficiency, namely the ratio between the total number of compact binaries (formed from a coeval population) that merge within an Hubble time  and the total initial mass of the stellar population with metallicity $Z$. The cosmological parameters used in eq.~\ref{eq:mrd} are taken  from \cite{planck2016}. The maximum formation redshift of progenitor binary stars is $z_{\rm max}=15$. For the cosmic star formation rate $\psi{}(z')$ we use the fitting formula from \citep{madau2017}, while for the evolution of metallicity we use equation~8 of \cite{santoliquido2021}, based on \cite{decia2018}. Metallicity evolution is extremely uncertain  but only has a mild effect on BNS mergers \citep{dominik2013,santoliquido2021}. Hence, this choice almost does not affect our results.
We obtain a local merger rate density $\mathcal{R}(z=0)=31$, 258 and  765  Gpc$^{-3}$ yr$^{-1}$, for the simulation set with $\alpha{}=1$, 3 and 7, respectively (Table~\ref{tab:GW}). These  values approximately bracket the 90\% credible interval estimated by the LIGO--Virgo--KAGRA collaboration in the GWTC-2.1 catalog \citep{2021arXiv210801045T}: $\mathcal{R}=286_{-237}^{+510}$  Gpc$^{-3}$ yr$^{-1}$. %\citep{abbottO3apopandrate}. 
This rate was inferred by the LIGO--Virgo-KAGRA collaboration assuming a BNS population uniform in component masses between 1.0 and 2.5 M$_\odot$. The GWTC-3 population paper \citep{2021arXiv211103634T} recently reported a larger interval $\mathcal{R}=13-1900$  Gpc$^{-3}$ yr$^{-1}$, which comes from the union of the 90\% credible intervals of several  population models, based on different assumptions for the BNS mass function.

Here, we only show  models with rate $\mathcal{R}(z=0)\in{}[31,\,{}765]$ Gpc$^{-3}$ yr$^{-1}$, because rates outside this range are extremely difficult to produce with astrophysically motivated models. Namely, rates lower than $\sim{30}$~Gpc$^{-3}$~yr$^{-1}$ can be achieved only with extremely large natal kicks (e.g., \citealt{santoliquido2021}). Rates higher than $\sim{800}$ Gpc$^{-3}$ yr$^{-1}$ are almost impossible to obtain with astrophysical models \citep{oshaughnessy2010,demink2015,mapelli2018,kruckow2018,vignagomez2018,neijssel2019,artale2019,eldridge2019,giacobbo2020,zevin2020,broekgaarden2021,mandel2021}, unless we allow Hertzsprung gap stars to survive a common envelope phase \citep{dominik2013,chruslinska2018}. 
%\barbara{I would add also the GWTC-2.1 rate: $\mathcal{R}=286_{-237}^{+510}$  Gpc$^{-3}$ yr$^{-1}$ \cite{2021arXiv210801045T} .Also, I would add a comment related to the GWTC-3 population paper, \cite{2021arXiv211103634T}} 
Hereafter, we refer to the simulation sets with $\alpha{}=1$, 3 and 7 as models A1, A3 and A7.

%\textcolor{magenta}{$\rightarrow$ Michela}

%\subsection{The cosmic merger rate density}

\section{The GW simulations}\label{sec:GW}
For each of the catalogs of BNSs described in Sec. \ref{sec:population}, we extracted 10$^5$ binaries and simulated the associated GW signals.
 To simulate the GW signal associated with each merger  we need the mass, sky position and spin of each system. We used the masses of the two NSs as reported in the catalogs. We assumed an isotropic and homogeneous distribution in space for our BNSs; we also assigned a random inclination of the orbital plane with respect to the line of sight $\theta_j$ to each BNS. As in \cite{2016JCAP...11..056P}, for simplicity we considered non-spinning systems, as the NS spin is expected to be small in compact binaries. In fact, only low-spin BNS have been observed through EM waves up to date: the most rapidly rotating pulsar found in a binary system, i.e. PSR J0737-3039A, has a period of $\sim$ 22.7 ms \citep{2003Natur.426..531B,2012PhRvD..86h4017B}, corresponding to a very low spin\footnote{$\chi$ is defined as $c\,{}J/G\,{}M^2$, where $c$ is the speed of light, $G$ is the gravitational constant and $J$ and $M$ are the angular momentum and the mass of the star, respectively.}: 
$\chi \sim$ 0.05. However, it is worth to mention that the fastest-spinning observed millisecond pulsar, i.e. PSR J1748-2446ad, has a lower period of $\sim$ 1 ms \citep{2006Sci...311.1901H} and then a higher spin: $\chi \sim$ 0.4; furthermore, even higher values are allowed by theoretical models.

For each merging BNS, we simulated the expected GW inspiral signals using the ``TaylorT4''  waveforms (see, e.g., \citealp{2009PhRvD..80h4043B}), that are constructed using post-Newtonian models accurate to the 3.5 order in phase and 1.5 order in amplitude. Then, we added the GW signal to the detector noise, assumed to be Gaussian. We considered a GW network composed of Advanced LIGO, Advanced Virgo and KAGRA, and used the sensitivity curves expected for O4 \citep{2020LRR....23....3A}. Specifically, we used the public sensitivities\footnote{https://dcc.ligo.org/LIGO-T2000012/public}: % available at \url{https://dcc.ligo.org/LIGO-T2000012/public}:
an intermediate sensitivity curve for KAGRA, corresponding to a BNS range of 80 Mpc, and the target sensitivity curve (the highest O4 sensitivity) for Advanced LIGO  and for Advanced Virgo, corresponding to a BNS range of 190 Mpc and 120 Mpc respectively. The data obtained in this way have been then analyzed with the matched filtering technique  \citep{wainstein63}. We consider two different scenarios to estimate the GW detection rates. In the first scenario, hereafter ``case a'' %Similarly to \cite{2016JCAP...11..056P} and \cite{2020LRR....23....3A}, 
we assume a source to be detected if it has a signal-to-noise ratio (SNR) larger than 4 in at least two detectors and a network SNR larger than 12, similarly to what has been done in \cite{2016JCAP...11..056P} and \cite{2020LRR....23....3A}. 
 We then consider another scenario (``case b'') in which a BNS merger is assumed to be detected if it has a network SNR larger than 8, even if observed with a single interferometer,  as done in \cite{2022ApJ...924...54P}. This last approach is representative of the public GW alerts sent during the third LIGO-Virgo-KAGRA observing run (O3); however, it is important to mention that a lower SNR threshold could imply a larger contamination by noise events, and that 30 \% of the O3 alerts have been retracted\footnote{https://gracedb.ligo.org/superevents/public/O3/}. In both scenarios we 
%We 
assume that each GW detector has an independent duty cycle of 70~\% and  consider an observing period of 1 year, that is the expected duration of O4 \citep{2020LRR....23....3A}. For the detected sources, we produced the associated 3-dimensional GW sky-maps with BAYESTAR, a rapid Bayesian position reconstruction code that computes source location using the output from the detection pipelines \citep{2014ApJ...795..105S,2016ApJ...829L..15S}. The matched filter pipeline and the sky localization have been simulated using the \texttt{bayestar-realize-coincs} and \texttt{bayestar-localize-coincs} tools from the \texttt{ligo.skymap} public library\footnote{\url{https://lscsoft.docs.ligo.org/ligo.skymap/}}.

The GW detection rates have been estimated by combining the BNS merger rate density and the percentage of detected BNS systems. 

\section{The associated GRB emission}\label{sec:GRB}
In order to evaluate the joint GW--EM detection, in what follows we assumed that all BNS mergers give rise to a S-GRB.

\subsection{The prompt emission}\label{sect:prompt}
We associated a set of simulated parameters describing the prompt emission of S-GRBs to each BNS merger: the rest-frame peak energy of the prompt emission spectrum $E_{\rm pk}$, the isotropic bolometric peak luminosity $L_{\rm iso}$, the redshift $z$ %, set equal to the one 
of the BNS merger, and the observer's viewing angle, that corresponds to the inclination of the BNS system $\theta_j$. 

Following the methodology presented in \citet{2016A&A...594A..84G}, hereafter G16, we assigned to each BNS merger a value of $E_{\rm pk}$ drawn randomly from a broken power law distribution (see G16, eq. (13), with the parameters reported in Table 1, case a) and a value of $L_{\rm iso}$ sampled from a lognormal distribution whose central value is given by the $E_{\rm pk}$--$L_{\rm iso}$ correlation (\citealt{2004ApJ...609..935Y}, written as in G16, eq. (14)) and $\sigma=0.2$ to account for the uncertainty in the parameters of the correlation. The average values for the $E_{\rm pk}$ and $L_{\rm iso}$ obtained with this procedure are: $\langle E_{\rm pk}\rangle = 700$ keV and $\langle L_{\rm iso}\rangle = 2\times 10^{52}$ erg~s$^{-1}$, and the minimum and maximum values allowed are: $E_{\rm pk}^{\rm min} = 0.1$ keV and $E_{\rm pk}^{\rm max} = 10^5$ keV, and $L_{\rm iso}^{\rm min} \sim 4\times 10^{45}$ erg s$^{-1}$ and $L_{\rm iso}^{\rm max} \sim 4\times 10^{55}$ erg s$^{-1}$.

We assumed as first approximation that the outflow of S-GRBs is a uniform jet, namely that the radiated energy and Lorentz factor are uniformly distributed inside a narrow jet with an opening angle $\theta_c$, and we explored two cases: $\theta_c=5^\circ$ and $10^\circ$ \citep[G16,][]{2015ApJ...815..102F}. Then, for S-GRBs associated to {\em on-axis} BNS mergers (i.e. BNS mergers with inclination $\theta_j<\theta_c$), we calculated the observed photon flux $P_{\rm pk}$ in the energy band $[E_{\rm min},E_{\rm max}]$ corresponding to a specific instrument as:
\begin{equation}
    P_{\rm pk}=\frac{(1+z) \int_{(1+z)E_{\rm min}}^{(1+z)E_{\rm max}}S(E)\,{}dE}{4\,{}\pi\,{} d_L^2} \, ,
\end{equation}
where $d_L$ is the luminosity distance at redshift z and $S(E)$ is a Band photon spectrum whose normalisation is such that:
\begin{equation}
    L_{\rm iso}=\int_{E_1}^{E_2} S(E)EdE\, ,
\end{equation}
with $[E_1,E_2]=[1,10]$ keV. For the Band spectrum, we assumed a constant value for the low and high photon spectral indices $\alpha_1=-0.6$ and $\alpha_2=-2.5$, respectively, representative of the spectral indices of the population of S-GRBs observed by \emph{Fermi}/GBM and well-fitted to a Band function \citep{2014ApJS..211...12G}. We keep these two parameters fixed after checking that our results are unaffected by sampling them from distributions centred around these values (see also G16).

We also explored a more reliable scenario by modelling the GRB outflow as a structured jet with two possible intrinsic structures for the radiated luminosity per jet unit solid angle and the Lorentz factor of the emitting material $\Gamma_\circ$ \citep{2015MNRAS.450.3549S,2019A&A...628A..18S}:
\begin{itemize}
    \item a power-law distribution:
    \begin{equation}
    \label{Etheta_pl}
     \frac{dL}{d\Omega}(\theta)=\frac{L_{\rm iso}/4\pi}{1+(\theta/5^\circ )^s}\, ,\quad
    \Gamma_\circ(\theta)=1+\frac{\Gamma_\circ-1}{1+(\theta/5^\circ )^p}\, .
    \end{equation}
    The power-law index s in the angular distribution of the radiated energy is allowed to vary within the range $3.5-5.5$, as estimated for GRB 170817A (see e.g. \citealt{2019Sci...363..968G} or \citealt{2018A&A...613L...1D}). For the Lorentz factor, we assumed that $\Gamma_\circ=250$ and $p=2$ \citep{2019Sci...363..968G};
    \item a gaussian distribution:
    \begin{equation}
    \label{Etheta_gauss}
     \frac{dL}{d\Omega}(\theta)=\frac{L_{\rm iso}}{4\pi}\,{\rm e}^{-(\theta/5^\circ)^2}\, ,\quad
    \Gamma_\circ(\theta)=1+(\Gamma_\circ-1)\,{\rm e}^{-(\theta/5^\circ)^2}\, ,
    \end{equation}
    with $\Gamma_\circ=250$ \citep{2019Sci...363..968G}.
\end{itemize}

The isotropic-equivalent luminosity observed at a viewing angle $\theta_j$ is then computed as \citep{2015MNRAS.450.3549S,2019A&A...628A..18S}:
\begin{equation}
\label{Liso_app}
    L_{\rm iso}(\theta_j)=\int \frac{\delta^3 (\theta,\phi,\theta_j)}{\Gamma_\circ(\theta)}\frac{dL}{d\Omega}(\theta)\,d\Omega\, ,
\end{equation}
where $\delta$ is the relativistic Doppler factor.

This allowed us to considered also the contribution to the detection rate by S-GRBs associated to moderately off-axis BNS mergers by calculating from eq. \ref{Liso_app} the isotropic-equivalent radiated luminosity for an observer at a viewing angle $\theta_j$ with $5^\circ< \theta_j<35^\circ$. Assuming that the $E_{\rm pk}$-$L_{\rm iso}$ correlation also holds for moderately off-axis S-GRBs, we calculated $E_{\rm pk}(\theta_j)$ from $L_{\rm iso}(\theta_j)$. While this assumption has been proved in simulations of the prompt emission of off-axis GRBs  (see, e.g., \citealt{2015MNRAS.450.3549S,2019A&A...628A..18S}), notably this is not the case for GRB 170817A. More detailed modelling of the prompt emission from off-axis S-GRBs is beyond the scope of the present work.

\subsection{The X-ray afterglow emission}\label{sec:afterglow}
Despite the number of high-energy satellites in operation with instrumentation dedicated to the detection of GRBs, it is possible that a S-GRB on axis or moderately off-axis is not detected during its prompt phase. This happens because even all-sky detectors (like \emph{Fermi}/GBM) have relatively limited duty-cycles due to observational constraints (see Appendix \ref{instr}). However, it is still possible to discover them by searching for their afterglows with dedicated facilities that are capable of tiling large regions in the sky and discover new transient sources in those regions. While for the X-ray afterglow emission we can rely on the robust and homogeneous sample collected with \emph{Swift}/XRT\footnote{$\sim 90\%$ of the S-GRBs detected by \emph{Swift}/BAT for which the satellite carried out a prompt slew has an afterglow detected by \emph{Swift}/XRT}, the sample of S-GRB afterglow light curves observed in the optical and radio bands is much more sparse, with less than $\sim 40\%$ and $\sim 10\%$ of the {\it Swift} S-GRBs having an optical and radio afterglow, respectively, detected \citep{2015ApJ...815..102F}. The optical and radio samples are therefore severely affected by selection effects. We thus evaluated how many BNS mergers leading to S-GRBs meet the optimal conditions for being discovered by facilities operating in the X-ray band. In what follows, we assume the strategy put in place by \emph{Swift}/XRT during the O3 run as a reference (\citealp{2016MNRAS.455.1522E,2016MNRAS.462.1591E}; for details see Appendix \ref{instr}).

The first requirement is that the skymap is a single region with a 90\% confidence area  $\leq{}50$ deg$^2$. This choice maximises the coverage of the whole region with a reasonable exposure time within $\sim$ 1 day \citep{2016MNRAS.462.1591E}, i.e. when the source is brighter. Then, in order to evaluate the detectability of the S-GRB X-ray emission at different times after the GW trigger, we took the sample presented in \citet{2014MNRAS.442.2342D} as reference for the population of on-axis S-GRBs, i.e. S-GRBs associated to BNS mergers with inclination $\theta_j<\theta_c$ with $\theta_c=5^\circ$ (see Sect. \ref{sect:prompt}). For each event of the sample we estimated the afterglow X-ray isotropic-equivalent luminosity in the 1--10 keV rest frame common energy band $L_{\rm X,rf}^{1-10}$ from the observed integral 0.3--10 keV unabsorbed flux $f_{\rm X}^{0.3-10}$ in the following way \citep[see][]{2014MNRAS.442.2342D}:  

\begin{equation} 
L_{\rm X,rf}^{1-10} = 4\pi d_L^2\, f_{\rm X}^{0.3-10}\frac{\left({\frac{10}{1+z}}\right)^{2-\Gamma}-\left({\frac{1}{1+z}}\right)^{2-\Gamma}}{{10}^{2-\Gamma}-{0.3}^{2-\Gamma}}\, ,
\label{kcorr_eq}
\end{equation}

where $\Gamma$ is the measured spectral index that we retrieved from the online {\it Swift Burst Analyser\footnote{\url{https://www.swift.ac.uk/burst_analyser}}} \citep{2009MNRAS.397.1177E,2010A&A...519A.102E}. The resulting X-ray afterglow light curves have been compared with the limiting luminosity that can be reached by {\it Swift}/XRT at different distances within the volume sampled in this work to calculate the fraction of on-axis BNS mergers that can be discovered within the first day or, in case of a non-detection during the first day, the fraction of mergers that are detectable in the following days with longer exposures (Figure \ref{fig:xray}; see Appendix \ref{instr}).

Under the assumption of a structured jet outflow (see Sect. \ref{sect:prompt}), we can also extend the possible detection to off-axis S-GRBs ($\theta_j>5^\circ$), by requiring that their X-ray emission peaks during the first $\sim$ 10 days after the merger in order to be effectively discovered during the monitoring of the error region. An estimate of the peak time as a function of the viewing angle can be obtained from: $t_{\rm pk}\sim 2.1\, E_{\rm k,52}^{1/3}n^{-1/3}((\theta_j-5^\circ)/10^\circ)^{8/3}$ days (see \citealp{2018ApJ...856L..18M}, from \citealp{2002ApJ...568..820G,2002ApJ...570L..61G}), where $E_{\rm k,52}$ is the GRB kinetic energy $E_{\rm k}$, in units of 10$^{52}$ erg, that can be obtained from $E_{\rm iso}$ via an efficiency factor: $E_{\rm k}=E_{\rm iso}/\eta$, and $n$ is the particle number density of the circumstellar medium. By assuming that $\langle E_{\rm iso}\rangle=\langle L_{\rm iso}\rangle/2\sim 10^{52}$ erg, $\eta =0.1$ and $n=10^{-3}$ cm$^{-3}$, we expect that the afterglow peaks at $\sim$ 6 days for $\theta_j\sim 10^\circ$. This simple estimate shows that the discovery of the X-ray counterpart of BNS mergers with $\theta_j>10^\circ$ with facilities as \emph{Swift}/XRT is highly unlikely. As a reference, in Figure \ref{fig:xray} we portray the X-ray light curve of GRB 170817A, that has an inclination of $\theta_j\sim 20^\circ$.

%\section{The GRB detectability}
%\barbara{Not sure this section is needed; eventually the Ref to this section should be added in last paragraph of introduction}

%\mg{Agreed}

\section{Results}\label{sec:results}
Tables \ref{tab:GW} and \ref{tab:GW2} show the expected number of GW detections in the 1-year O4 science run for the first and the second scenario respectively. When the first scenario is considered, the   number of detections ranges from 1 to 13 events per year, depending on the theoretical model used to simulate the population of BNS mergers. These numbers are within the range predicted in \cite{2020LRR....23....3A}, where a much larger range of values for the local BNS merger rate density has been considered. When a lower SNR threshold is considered and observations with a single interferometer are included, the number of detections ranges instead from 5 to 61 events per year: there is therefore an increase  by a factor of $\sim$ 5. These numbers are consistent with the rate of BNS detections predicted for O4 by \cite{2022ApJ...924...54P}, taking into account the difference in the assumed range of values for the BNS merger rate density.%These numbers are consistent with the ones reported in \cite{2021arXiv210807277P}}

\begin{table*}
%\begin{sidewaystable*}
%\begin{landscape}
    \centering
    \begin{tabular}{c|c|c|cc|cc|cc|cc}
        \hline
         Model & $\mathcal{R}(0)$ & GW &  \multicolumn{8}{|c|}{GW+EM (prompt)} \\
         \hline
          & & & \multicolumn{2}{|c|}{Swift/BAT} & \multicolumn{2}{|c|}{Fermi/GBM} & \multicolumn{2}{|c|}{INTEGRAL/IBIS} & \multicolumn{2}{|c|}{SVOM/ECLAIRs} \\
          & & & uniform & structured & uniform & structured & uniform & structured & uniform & structured \\
          & Gpc$^{-3}$yr$^{-1}$& yr$^{-1}$& yr$^{-1}$& yr$^{-1}$& yr$^{-1}$& yr$^{-1}$& yr$^{-1}$& yr$^{-1}$& yr$^{-1}$& yr$^{-1}$ \\ 

         \hline
          A1 & 31 & 1 &0.0006 (0.0023) &0.014-0.020 &0.003 (0.013) &0.070-0.11 &0.0001 (0.0004) &0.0024-0.0035 &0.0005 (0.0019) &0.013-0.017 \\
          A3 & 258 & 5 &0.003 (0.01) &0.07-0.10 &0.017 (0.068) &0.35-0.54 &0.0005 (0.002) &0.01-0.02 &0.002 (0.01) &0.06-0.08 \\
          A7 & 765 & 13 &0.008 (0.031) &0.18-0.26 &0.045 (0.18) &0.91-1.42 &0.001 (0.005) &0.031-0.046 &0.006 (0.025) &0.17-0.22 \\
          \hline
    \end{tabular}
    \caption{Expected rates of GW and joint GW+EM detection of BNS mergers and the associated GRB prompt emission for the three BNS models considered in this work for the most conservative scenario for the GW detection (case a); the local BNS merger rate density $\mathcal{R}(0)$ is also reported. The rates for the joint GW+EM detection are computed for a uniform jet with aperture $\theta_c=5^\circ$ ($\theta_c=10^\circ$) and for a structured jet. In this case, the range reported corresponds to the minimum and maximum rate obtained for the different structures of the jet assumed.}
    \label{tab:GW}
%\end{landscape}
%\end{sidewaystable*}
\end{table*}

%\begin{table*}
%    \centering
%    \begin{tabular}{c|c|c|cc|cc|cc|cc}
%        \hline
%         Model & $\mathcal{R}(0)$ & GW &  \multicolumn{8}{|c|}{GW+EM (prompt)} \\
%         \hline
%          & & & \multicolumn{2}{|c|}{Swift/BAT} & \multicolumn{2}{|c|}{SVOM/ECLAIRs} & %\multicolumn{2}{|c|}{Fermi/GBM} & \multicolumn{2}{|c|}{INTEGRAL/IBIS} \\
%          & & & uniform & structured & uniform & structured & uniform & structured & uniform & structured \\
%          & Gpc$^{-3}$ yr$^{-1}$& yr$^{-1}$& yr$^{-1}$& yr$^{-1}$& yr$^{-1}$& yr$^{-1}$& yr$^{-1}$& yr$^{-1}$& %yr$^{-1}$& yr$^{-1}$ \\ 
%    \hline
%    A1 & 31 & 5 &0.002 (0.010) &0.05-0.08 (0.06) &0.002 (0.008) &0.05-0.07 (0.05) &0.014 (0.060) &0.27-0.46 (0.28) %&0.0005 (0.002) &0.009-0.014 (0.010) \\
%    A3 & 258 & 22 &0.01 (0.04) &0.24-0.37 (0.25) &0.009 (0.04) &0.22-0.32 (0.24) &0.06 (0.26) &1.17-2.00 (1.24) %&0.002 (0.008) &0.04-0.06 (0.04) \\
%    A7 & 765 & 61 &0.03 (0.12) &0.67-1.05 (0.71) &0.02 (0.10) &0.63-0.90 (0.66) &0.18 (0.74) &3.28-5.65 (3.44) %&0.006 (0.02) &0.11-0.18 (0.12) \\
%     \hline
%   \end{tabular}
%    \caption{Same as Table \ref{tab:GW}, but assuming a lower network SNR threshold of 8 for the GW detections and %including the GW signals detected with a single interferometer.}
%    \label{tab:GW2}
%\end{table*}

\begin{table*}
    \centering
    \begin{tabular}{c|c|c|cc|cc|cc|cc}
        \hline
         Model & $\mathcal{R}(0)$ & GW &  \multicolumn{8}{|c|}{GW+EM (prompt)} \\
         \hline
          & & & \multicolumn{2}{|c|}{Swift/BAT} & \multicolumn{2}{|c|}{Fermi/GBM} & \multicolumn{2}{|c|}{INTEGRAL/IBIS} & \multicolumn{2}{|c|}{SVOM/ECLAIRs} \\
          & & & uniform & structured & uniform & structured & uniform & structured & uniform & structured \\
          &Gpc$^{-3}$yr$^{-1}$ & yr$^{-1}$& yr$^{-1}$& yr$^{-1}$& yr$^{-1}$& yr$^{-1}$& yr$^{-1}$& yr$^{-1}$& yr$^{-1}$& yr$^{-1}$ \\ 
    \hline
    A1 & 31 & 5 &0.002 (0.01) &0.05-0.08 &0.014 (0.06) &0.27-0.46 &0.0005 (0.002) &0.009-0.014 &0.002 (0.008) &0.05-0.07 \\
    A3 & 258 & 22 &0.01 (0.04) &0.24-0.37 &0.06 (0.26) &1.17-2.00 &0.002 (0.008) &0.04-0.06 &0.009 (0.04) &0.22-0.32 \\
    A7 & 765 & 61 &0.03 (0.12) &0.67-1.05 &0.18 (0.74) &3.28-5.65 &0.006 (0.02) &0.11-0.18  &0.02 (0.10) &0.63-0.90 \\
     \hline
   \end{tabular}
    \caption{Same as Table \ref{tab:GW}, but assuming a lower network SNR threshold of 8 for the GW detections and including the GW signals detected with a single interferometer (case b).}
    \label{tab:GW2}
\end{table*}

\begin{table*}
    \centering
    \begin{tabular}{c|cc|cc|cc|cc}
        \hline
         Model &  \multicolumn{8}{|c|}{GW+EM (prompt)} \\
         \hline
          & \multicolumn{2}{|c|}{Swift/BAT} & \multicolumn{2}{|c|}{Fermi/GBM} & \multicolumn{2}{|c|}{INTEGRAL/IBIS} & \multicolumn{2}{|c|}{SVOM/ECLAIRs} \\
          & case a & case b & case a & case b & case a & case b & case a & case b \\
          & yr$^{-1}$& yr$^{-1}$& yr$^{-1}$& yr$^{-1}$& yr$^{-1}$& yr$^{-1}$& yr$^{-1}$& yr$^{-1}$ \\ 
    \hline
    A1 &0.015 &0.06 &0.073 &0.28 &0.0025 &0.01 &0.013 &0.05 \\
    A3 &0.017 &0.25 &0.37 &1.24 &0.010 &0.04 &0.07 &0.24 \\
    A7 &0.19 &0.71 &0.96 &3.44 &0.032 &0.12 &0.17 &0.66 \\
     \hline
   \end{tabular}
    \caption{Expected rates of joint GW+EM detection of BNS mergers and the associated GRB prompt emission for the three BNS models and the two scenarios for the GW detection considered in this work (case a and b); these rates are computed assuming an intrinsic structure of the jet with a gaussian distribution.}
    \label{tab:gauss}
\end{table*}

Then, we evaluated the fraction of mergers that give rise to a S-GRB whose prompt emission is detectable by the main GRB facilities. A description of these facilities and the assumptions adopted for the energy band, detection threshold, field of view (fov) and duty cycle are reported in Appendix \ref{instr}.
Tables \ref{tab:GW} and \ref{tab:GW2} show our results for the two GW detection scenarios (case a and b), respectively. In the uniform jet scenario, the EM-GW joint detection rate is highly suppressed by the geometry of the jet. In fact the fraction of GW events with inclination lower than $5^\circ$ ($10^\circ$) is $0.6\%$ ($2.5\%$).

Similar results were found in G16 with the same population. In G16, the detection rate of S-GRBs within a radius of 450 Mpc is $\sim 0.08$ yr$^{-1}$, without any a-priori assumption on the jet opening angle and BNS merger rate. Since they do not make specific assumptions on the detection instrument, we compared this figure with the rate obtained for \emph{Fermi}/GBM corrected for its limited duty cycle. With all these caveats, we find a fair agreement with our rates for the uniform jet case, with closer values for the model A3. Since the rate in G16 has been derived using all the available observer-frame constraints of the large population of \emph{Fermi} S-GRBs and the rest-frame properties of a complete sample of S-GRBs detected by \emph{Swift}, the consistency we find with the predictions derived by Ghirlanda et al. 2016 ensures that our detection rates are consistent with the observed rates for \emph{Swift} and \emph{Fermi}.

Indeed, GRB 170817A showed that the outflow of S-GRBs is likely structured, allowing for a possible detection of their prompt emission even when they are observed at larger viewing angles. Since our knowledge of the jet structure is limited to one case, we considered different possibilities for the jet structure to cover a broad range of possibilities. The minimum and maximum detection rates correspond both to the power-law intrinsic structure with the steeper and shallower indices, respectively. The Gaussian intrinsic structure gives detection rates within these ranges (see Table \ref{tab:gauss}). The structured S-GRB outflow increases the detection rate by a factor of $\sim 20-40$ with respect to the case of an uniform jet with an aperture $\theta_c=5^\circ$ (a factor $\sim 5-8$ for $\theta_c=10^\circ$). For the structured jet case, the fraction of GW events with joint EM detection is $\sim 1-2\%$ for \emph{Swift}/BAT, $7-11\%$ for \emph{Fermi}/GBM and $0.2-0.3\%$ for INTEGRAL/IBIS, depending on the jet structure. 
These figures translate to a joint GW and EM detection rate that is $\sim 1$ yr$^{-1}$ for \emph{Fermi}/GBM in the most favourable case (model A7) when we consider the most conservative scenario (case a, see Table \ref{tab:GW}). Although obtained under different assumptions, the expected rate of joint GW and GRB detection with \emph{Fermi}/GBM is consistent with the estimate reported in \citep{2021arXiv211103608T}. %, (1.04$^{+0.26}_{-0.27}$ yr$^{-1}$).} 
Perspectives for a joint GW-EM detection for the prompt emission are more promising allowing for a lower SNR threshold and less interferometers for the GW detection (case b, see Table \ref{tab:GW2}). In this scenario, we expect several detections per year for \emph{Fermi}/GBM, and up to one detection per year for \emph{Swift}/BAT for model A7.

\begin{figure}
    \centering
    \includegraphics[width=0.5\textwidth]{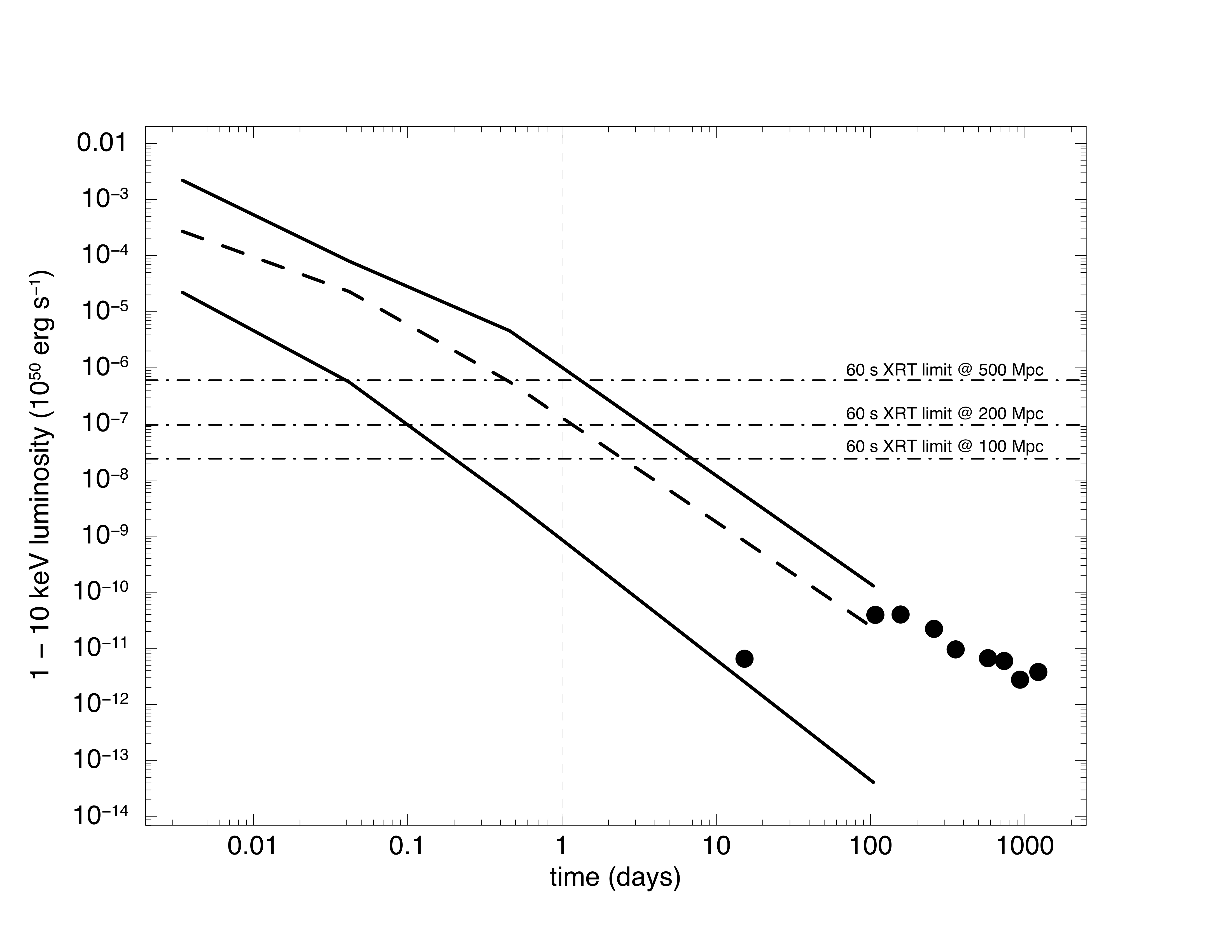}
    \includegraphics[width=0.5\textwidth]{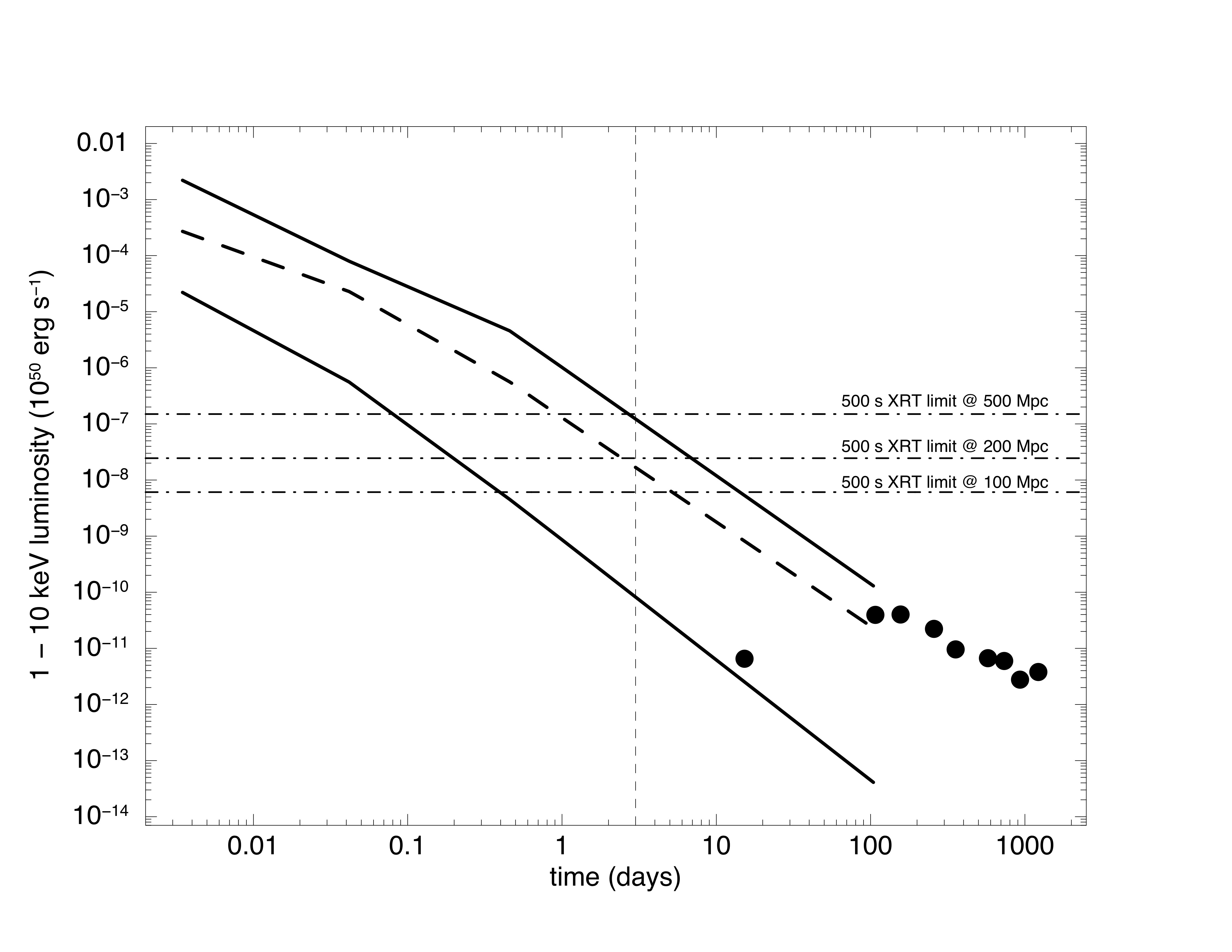}
    \caption{The distribution of S-GRB light curves of the S-BAT4 sample \citep{2014MNRAS.442.2342D}. The X-ray luminosities were computed for each GRB in the common rest-frame 1–10 keV energy band following the procedure described in Section 4.2. The dashed line shows the median behaviour; the two solid lines represent the 25 and 75 percentiles. The X-ray afterglow of GW\,170817 / GRB\,170817 is shown with black points for reference (adapted from \citealt{2021arXiv210402070H}). The dashed-dotted lines represent the depths that can be reached with {\it Swift/XRT} with the current strategy for GW follow-up.}
    \label{fig:xray}
\end{figure}

We also considered the possibility to discover the EM counterpart of the GW-detected events in X-rays with instruments like \emph{Swift}/XRT. Figure \ref{fig:xray} portrays the distribution of the X-ray luminosity of on-axis S-GRBs compared with the limiting luminosity for \emph{Swift}/XRT at 100, 200 and 500 Mpc (see Appendix \ref{instr}). About 60\% (50\%) of S-GRBs at 100 Mpc (200 Mpc) would be detectable by \emph{Swift}/XRT 1 day after the merger with an exposure of 60 s, while $\sim$ 55\% (45\%) would be still detectable after 3 days in case of a revisit with an exposure of 500 s. This percentage decreases if we move our horizon to 500 Mpc ($\sim$ 30\% at 1 day and 60 s exposure; $\sim 25\%$ at 3 days and 500 s exposure). We thus evaluated the number of GW-detected BNS mergers within these horizons (100, 200 and 500 Mpc) with optimal characteristics for being discovered by \emph{Swift}/XRT within the first few days, namely being on-axis ($\theta_j<5^\circ$) and  with a good localization (90\% confidence region $< 50$ deg$^2$). In the most conservative scenario (case a), this is $\sim$0.5\% of the GW-detected BNS mergers in the same volume.
%, that translates into a total number of 0.003, 0.012, 0.035 events per year within 100 Mpc (0.004, 0.02, 0.05 events per year within 200 Mpc) for the three BNS models considered in this work.  
In the less conservative scenario (case b) the expected rate increases, but less than for the GW detection rate since the detection with lower SNR and less detector provides a less accurate localization. Convolving these rates with the probability that the X-ray luminosity is above the flux limit at that distance, as described above, we can estimate the rate of X-ray counterparts associated to S-GRBs of GW-detected BNS mergers observable by \emph{Swift}/XRT within different volumes during the first day or the following days of observation (see Table~\ref{tab_aft}). If we consider mergers with $\theta_j<10^\circ$, namely on-axis S-GRBs with a wider core or moderately off-axis S-GRBs that peak within a few days after the merger, rates rise up to about a factor of four. In all cases, the rate of S-GRB associated to detectable GW events remains very low, due to the limited volume sampled by these searches and by the requirement to observe the S-GRB nearly on-axis.

\begin{table*}
    \centering
    \begin{tabular}{c|ccc|ccc}
    \hline
        Model & \multicolumn{3}{c}{GW+EM (X-ray afterglow), case a} & \multicolumn{3}{c}{GW+EM (X-ray afterglow), case b}\\
        \hline
         & < 100 Mpc  & 100-200 Mpc & 200-500 Mpc& < 100 Mpc  & 100-200 Mpc & 200-500 Mpc \\
         & yr$^{-1}$ & yr$^{-1}$ & yr$^{-1}$ & yr$^{-1}$ & yr$^{-1}$ & yr$^{-1}$ \\
        \hline
        A1 & 0.0015-0.0026 & 0.0007-0.0014 & 0.0002-0.0006 & 0.005-0.008 & 0.002-0.004 & 0.0007-0.0024 \\
        A3 & 0.007-0.012 & 0.003-0.007 & 0.001-0.003 & 0.019-0.032 & 0.010-0.019 & 0.004-0.013 \\
        A7 & 0.021-0.035 & 0.009-0.017 & 0.002-0.006 & 0.098-0.059 & 0.025-0.050 & 0.008-0.028 \\
        \hline
        A1 & 0.0014-0.0017 & 0.0006-0.0009 & 0.0002-0.0003 & 0.004-0.005 & 0.0018-0.0025 & 0.0006-0.0012 \\
        A3 &  0.007-0.010 & 0.003-0.004 & 0.0008-0.002 &  0.018-0.021 & 0.009-0.011 & 0.003-0.006 \\
        A7 &  0.019-0.023 & 0.008-0.010 & 0.001-0.003 & 0.054-0.064 & 0.022-0.030 & 0.007-0.014 \\
        \hline
    \end{tabular}
    \caption{Expected rates of X-ray counterparts of GW-detected BNS mergers discoverable by \emph{Swift}/XRT within different volumes (<100 Mpc, between 100 and 200 Mpc and between 200 and 500 Mpc) during the first day of observation with 60 s exposure (upper panel), and between the first and the third day with 500 s exposure (lower panel), for the three BNS models considered in this work and for the two possible scenarios for the GW detection (case a and b, respectively). The values reported are obtained considering GW-detected BNS mergers with $\theta_j<5^\circ$ and 90\% confidence region $< 50$ deg$^2$ convolved with the probability that the X-ray luminosity is above the flux limit at the corresponding distance (60\%, 50\% and 30\% with 60 s exposure, and 55\%, 45\% and 25\% with 500 s exposure, respectively; see Sect. \ref{sec:afterglow}). If we consider mergers with $\theta_j<10^\circ$, these rates rise up to about a factor of four.}
    \label{tab_aft}
\end{table*}

\section{Discussion and Conclusions}\label{sec:conclusions}

We presented a comprehensive study on the expectations for joint GW and EM detection of BNS mergers in the next observing run of Advanced LIGO, Advanced Virgo and KAGRA (O4), considering different EM facilities: \emph{Fermi}, \emph{Swift}, INTEGRAL and SVOM.

Depending on the population synthesis model considered, we expect to have from 1 up to 13 BNS merger detections per year when a conservative approach is considered (case a). The number of detections per year increases by a factor of $\sim$ 5 when we include the possibility to detect BNS mergers with a single interferometer and with a reduced network SNR threshold (case b, as done in  \citealp{2022ApJ...924...54P}). %These estimates are conservatives, since they do not consider the possibility to detect BNS mergers with a single interferometer or with a reduced network SNR threshold (see \citealp{2022ApJ...924...54P}). 
It is also worth to mention that the sensitivity of KAGRA during O4 could be lower than the one considered in this work (see \url{https://www.ligo.org/scientists/GWEMalerts.php}). However, this is not expected to affect our results in a significant way, since the main contribution to the GW detection rate comes from the more sensitive Advanced LIGO and Advanced Virgo detectors. For instance, if we assume that the KAGRA sensitivity is of the order of 1 Mpc for the whole duration of O4, the decrease in the GW detection rate is less than 10 $\%$.% \textbf{for both the scenarios  considered in this work}.

Concerning the joint detection of the GW signal and the S-GRB prompt emission, we show that during O4 this is not only possible, but also likely, depending on the jet structure and on the BNS rate. The best perspectives come from \emph{Fermi}/GBM that will likely detect one event, and possibly several events, during O4. The other instruments are limited by their fov. However, \emph{Swift}/BAT will be capable of detecting up to one event per year, with the major advantage of providing accurate localization and a rapid follow-up in X-rays (see Sec. \ref{sec:afterglow}).

%In this respect, the synergy between Swift and SVOM will be crucial also for the follow-up of SVOM-triggered events, since \emph{Swift}/XRT is more sensitive than SVOM/MXT and can: i) follow X-ray candidates discovered by SVOM/MXT for longer periods; ii) discover X-ray counterparts of SVOM/ECLAIRs triggers that are too faint or too off-axis to be detected by SVOM/MXT within the first hours.

The perspectives for the joint GW-EM detections are even more encouraging than what derived in the present work based on the \emph{standard} detection of S-GRBs, since there are strategies specifically tailored to increase the sensitivity of the instruments to the counterparts of GW events, as targeted searches for sub-threshold events \citep{2019arXiv190312597G}. Swift/BAT has implemented the Gamma-ray Urgent Archiver for Novel Opportunities (GUANO\footnote{\url{https://www.swift.psu.edu/guano/}}, see \citealt{2020ApJ...900...35T}), a pipeline designed for targeted recovery of BAT event-by-event data around the times of compelling astrophysical events to enable sensitive targeted searches as NITRATES (Non-Imaging Transient Reconstruction And TEmporal Search, \citealp{2021arXiv211101769D}) that boosts the discovery rate of GRB170817A-like events in BAT by a factor of at least $3-4$x. In addition, there are other instruments that can support those considered in this work for a more efficient discovery of S-GRBs. INTEGRAL/IBIS is limited by its fov, but S-GRBs could be also detected by the Anti-Coincidence Shield (ACS) of the SPI spectrometer\footnote{\url{https://www.cosmos.esa.int/web/integral/instruments-spi}} on-board the INTEGRAL spacecraft. The SPI-ACS works as a nearly omni-directional gamma-ray burst detector above $\sim$80 keV, but it lacks spatial and spectral information.

SVOM will be a further asset for the discovery of short GRBs when it will join the O4 run (its launch is expected for mid-2023). SVOM/ECLAIRs have similar performances than \emph{Swift}/BAT for the detection of S-GRBs (see tables \ref{tab:GW} and \ref{tab:GW2}) and, since they are not likely monitoring the same region of the sky, together they will almost double the possibility to have a GW event detected during the prompt emission when SVOM will be operational. Besides, SVOM will also be equipped with a Gamma-Ray burst Monitor (GRM) with a larger fov and a higher high-energy threshold (5 MeV) than ECLAIRs that will improve the detection of S-GRBs, although with poor localization (see e.g. \citealt{2016arXiv161006892W,2017ExA....44..113B}).

%\barbara{Sotto ho aggiunto un paragrafo con il confronto con un paio di lavori; non ho messo il confronto \cite{2020MNRAS.493.1633S} con perchè li includono anche i sub-thresholds nei detection rates.}

In the last years other authors investigated the prospects for joint detections of GWs and S-GRBs with different approaches and under different assumptions, that led to different estimates of the detection rates. For instance, \cite{2019ApJ...881L..40S} predicted a joint GW/S-GRB detection rate of 1.83, 0.388 and 0.668 yr$^{-1}$ for \emph{Fermi}/GBM, \emph{Swift}/BAT and SVOM/ECLAIRS respectively; these estimates are based on the assumption of a local BNS merger rate density of 1540$^{+3200}_{-1220}$ Gpc$^{-3}$ yr$^{-1}$, a network SNR GW threshold of 16  and of a universal jet profile (the one inferred for GRB 170817A) for all the S-GRBs associated with the BNS Mergers. Assuming the same local BNS merger rate density but a lower SNR GW threshold of 8, \cite{2019MNRAS.485.1435H} predicted instead a joint GW/S-GRB detection rate of 1.23$^{+2.55}_{-0.97}$ yr$^{-1}$ for \emph{Fermi}/GBM. This work differentiates from previous studies for several key aspects. We investigated the joint S-GRB and GW observations by combining accurate population synthesis modeling  with pipelines specifically developed to provide GW detections and low-latency GW sky localization: we use theoretically motivated BNS merger rate densities, each one corresponding to a specific physical model and we mimic the real GW data analysis. Furthermore, in the GRB modeling we do not rely on the properties of GRB 170817A only, but we simulated the whole S-GRB population starting from reliable assumptions (the luminosity function derived in G16 using all the available observer-frame constraints of the large population of Fermi S-GRBs and the rest-frame properties of a complete sample of S-GRBs detected by Swift, and the $E_{\rm pk}-L_{\rm iso}$ correlation) and we explored different possibilities for the jet structure, computing the apparent structure as introduced in \citet{2015MNRAS.450.3549S}.

The perspectives to observe X-ray counterparts to GW events are not very promising, mainly because we can detect only events pointing towards us ($\theta_j<10^\circ$). However, the improved localization expected in O4 for nearby events [we will have 52\% (45\%) of events within 200 Mpc with a 90\% credible region (c.r.) $< 10$ deg$^2$, and 42\% (37\%) with a 90\% c.r. $< 5$ deg$^2$ for case a (case b)] will permit a faster coverage of the region, catching the candidate when it is brighter.

In this work, we did not explore the contribution to the detection rate of galaxy-targeted searches of EM counterparts. While this approach has revealed itself successful in the case of GW170817, for more distant events this could not be the optimal observational strategy: in fact, current galaxy catalogs are complete only up to a distance of a few tens of Mpc, loosing more than half of the brightest galaxies beyond $\sim 250$ Mpc (see, e.g., \citealp{2016MNRAS.462.1591E,2021arXiv211006184D}).

Besides S-GRBs, a key signature of a BNS (and possibly NSBH) binary merger is the production of a kilonova whose emission is characterised by two main components, namely the blue and red kilonova \citep{2019LRR....23....1M}. The red kilonova component emission is expected to be nearly-isotropic and to peak in the optical/NIR bands, while the blue kilonova (whose emission peaks in the UV/optical band) might not be present in all BNS mergers \citep{2015MNRAS.446.1115M} and, even if present, its emission is expected to be angle-dependent (i.e. the brightness can depend on the line of sight; \citealt{2017Natur.551...80K}). %In a recent work, 
Recently, \cite{2021arXiv211101945A} highlighted that realistic predictions on joint GW--kilonova detection should conservatively built around the capability to detect the red kilonova component. In their work, the authors show that a firm detection of red kilonovae during O4 can be obtained only within $\sim$ 100 Mpc, while for joint GW--kilonova detection beyond such horizon we will have to wait for O5, when the Vera Rubin Observatory will be operational. From our study, we estimate that during O4 the rate of GW-detected BNS mergers within 100 Mpc is 0.6--8 yr$^{-1}$ for case a and 2--23 yr$^{-1}$ for case b, depending on the model considered, and all these events are possible targets for the search of the associated red kilonova.

%\textbf{This work has shown there is a high probability to observe  at least another multi-messenger event  during O4, provided that EM facilities such as \emph{Fermi}/GBM and \emph{Swift}/BAT will be operating. This highlights the crucial role of small and medium size space missions to discover and characterize EM counterparts to GW events, and therefore  the need of a dedicated time-domain program of missions to sustain the necessary suite of space-based EM capabilities required to look for EM counterparts to GW events (see, e.g., ``Pathways to Discovery in Astronomy and Astrophysics for the 2020s'', the National Academies of Science, Engineering and Medicine's latest decadal survey\footnote{\url{https://nap.edu/resource/26141/interactive/}}).  } 

All the results presented in this paper take into account the current theoretical uncertainties on the BNS merger rate density, as well as our poor knowledge of the GRB population and of the jet structure. A direct comparison of these results with future joint GW-EM observations will narrow down most of these unknowns, shedding light on the physics of compact objects and on the association between S-GRBs and BNS system, as well as on the GRB jet geometry. Indeed, there is a realistic probability to observe at least another multi-messenger event during O4, provided that EM facilities such as \emph{Fermi}/GBM and \emph{Swift}/BAT will be operating. As demonstrated by the case of GW 170817/GRB 170817A, such an achievement would lead to a significant scientific advancement in the newly-born field of multi-messenger astronomy. As highlighted in this work, this will be possible only with dedicated time-domain programs of space and ground-based facilities with the necessary suite of capabilities required to detect and follow-up EM counterparts to GW events (see, e.g., “Pathways to Discovery in Astronomy and Astrophysics for the 2020s”, the National Academies of Science, Engineering and Medicine’s latest decadal survey\footnote{\url{https://nap.edu/resource/26141/interactive/}}).

\section*{Acknowledgements}
We are grateful to the referee for the very valuable comments and suggestions. We thank Sergio Campana for useful discussions. MM and FS acknowledge financial support from the European Research Council for the ERC Consolidator grant DEMOBLACK, under contract no. 770017. BP acknowledges financial support from the ESCAPE project with grant no. GA:824064. This work makes use of the \texttt{ligo.skymap} scientific software package (\url{https://lscsoft.docs.ligo.org/ligo.skymap/}). 
%%%%%%%%%%%%%%%%%%%%%%%%%%%%%%%%%%%%%%%%%%%%%%%%%%
\section*{Data Availability}
The data underlying this article will be shared on reasonable request to the corresponding author.

\bibliographystyle{mnras}
\bibliography{Bibliography}

%%%%%%%%%%%%%%%%%%%%%%%%%%%%%%%%%%%%%%%%%%%%%%%%%%

%%%%%%%%%%%%%%%%% APPENDICES %%%%%%%%%%%%%%%%%%%%%

\appendix

\section{Instruments}\label{instr}

\textbf{\emph{Swift}/BAT and XRT:} {\it Swift} \citep{2004ApJ...611.1005G} is a US/UK/Italian multi-wavelength space observatory dedicated to the study of GRB science. Launched in November 2004, it is equipped with three instruments, a wide-field hard X-ray burst detection telescope (Burst Alert Telescope— BAT; operating in the 15-150 keV band) with narrow-field X-ray (X-Ray Telescope—XRT; operating in the 0.3-10 keV band) and ultraviolet-optical telescopes (UV Optical Telescope–UVOT; operating in the 170-600 nm spectral range). These three instruments work together to provide imaging, timing and spectroscopic observations of GRBs and afterglows in the gamma-ray, X-ray, ultraviolet, and optical wavebands. For {\it Swift}/BAT, in this work we adopted a duty cycle of 90\%, a fraction of the sky covered of fov/4$\pi$ with f.o.v.=1.4 sr and a flux limit for the detection $P_{\rm lim}=0.6$ ph/s/cm$^2$ in the energy band 15--150 keV. To estimate the {\it Swift}/XRT performances we took as reference the strategy outlined in \citep{2016MNRAS.462.1591E}. According to such a strategy, each field of the GW skymap is visited at least twice, with a first visit carried out as soon as possible with exposure time of 60 $s$ and a second visit carried out with a longer exposure (500 $s$). The pointing capabilities of {\it Swift} enable to cover $\sim 50$ sq deg per day. For our purposes, we considered a detection when 5 (10) counts are detected with a 60 $s$ (500 $s$) exposure. Using WebPIMMS\footnote{\url{https://heasarc.gsfc.nasa.gov/cgi-bin/Tools/w3pimms/w3pimms.pl}} we converted these count/rates into fluxes in the common 1 - 10 keV energy band (see Sect. 4.2) assuming a typical GRB spectrum (an absorbed power-law with photon index 2) and a Galactic $N_H=3 \time 10^{20} \, cm^{-2}$ and then in the corresponding limiting 1--10 keV luminosity at 100, 200 and 500 Mpc.  

\noindent \textbf{\emph{Fermi}/GBM:} 
The Fermi Gamma-Ray Space Telescope was launched on 2008 June. The payload comprises two science instruments, the Large Area Telescope (LAT) and the Gamma-Ray Burst Monitor (GBM). The GBM \citep{2009ApJ...702..791M} is a nearly all-sky monitor, capable to trigger GRBs between $\sim 8$ keV and $\sim 40$ MeV and to study the spectrum and time history of the GRB prompt emission. In this work we assumed that the whole sky is monitored with a duty cycle of 60\%\footnote{\url{https://fermi.gsfc.nasa.gov/ssc/observations/types/grbs/}}, and a flux limit for the detection $P_{\rm lim}=0.27$ ph/s/cm$^2$ in the energy band 50--300 keV. This value corresponds to the peak flux above which lie 95\% of the S-GRBs from the Fermi/GBM catalog (\citealt{2020ApJ...893...46V}, see also \footnote{\url{https://fermi.gsfc.nasa.gov/science/instruments/table1-2.html}}).

\noindent \textbf{INTEGRAL/IBIS:} INTEGRAL is a satellite of the European Space Agency, launched in 2002. The instrument used for the GRB detection is IBIS, which provides images in the 15 keV-1 MeV range over a large field of view of 30x30 degrees using the coded mask technique \citep{2021NewAR..9301629K}. No on-board triggering system is present on INTEGRAL. The search for GRBs is done on ground, where the data arrive with an average delay of $\sim$10 s, in nominal conditions. For this purpose, the IBAS system has been implemented for the real time detection of GRBs and the rapid distribution of their coordinates \citep{2003A&A...411L.291M}. In this work we adopted a duty cycle of 85\% \citep{2021NewAR..9301629K}, a fraction of the sky covered of f.o.v./4$\pi$ with f.o.v.=30$^\circ \times$ 30$^\circ$ and a flux limit for the detection $P_{\rm lim}=0.15$ ct/s/cm$^2$ in the energy band 50--300 keV\footnote{\url{http://ibas.iasf-milano.inaf.it/}}.

\noindent \textbf{SVOM/ECLAIRs:} SVOM (Space-based multiband astronomical Variable Objects Monitor, \citealp{2016arXiv161006892W}) is a sino-french mission that is dedicated to Gamma-Ray Burst (GRB) science, expected to be launched in mid 2023. The mission includes four space-based and three ground-based instruments for a complete monitoring of the GRB emission, from the prompt to the late phases of the afterglow. The telecope ECLAIRs is a coded mask instrument with a f.o.v. of 2 sr, capable of triggering and locating GRBs with a precision of less than 12' in the 4--120 keV energy band. In this work we adopted a duty cycle of 50\%, due to the occultation of the Earth \citep{2016arXiv161006892W}, a fraction of the sky covered of fov/4$\pi$ with fov=2 sr and a flux limit for the detection $P_{\rm lim}=1.79$ ph/s/cm$^2$ in the energy band 4--120 keV \citep{2021A&A...645A..18D}.
%%%%%%%%%%%%%%%%%%%%%%%%%%%%%%%%%%%%%%%%%%%%%%%%%%

%\section{The gaussian structured jet}\label{gauss_sect}

% Don't change these lines
\bsp	% typesetting comment
\label{lastpage}
\end{document}